\documentclass{aa}
\usepackage{txfonts}
\usepackage{graphicx} 

\newcommand*{\emes}{\mathcal{M}}    
\newcommand*{\eturn}{{E_\mathrm{turn}}} 
\newcommand*{\eref}{{E_0}} 
\newcommand*{\ftot}{F_\mathrm{tot}}
\newcommand*{\flux}[1]{F_{\!{#1}}}

\begin{document}
%

\title {The spectral evolution of impulsive solar X-ray flares}

\titlerunning {Spectral evolution of solar X-ray flares}

\author{Paolo C. Grigis \and Arnold O. Benz}

\offprints{Paolo C. Grigis,
\email{pgrigis@astro.phys.ethz.ch}}

\institute{Institute for Astronomy, ETH Z\"urich, CH-8092 Z\"urich}

\date{Received 28 May 2004 / Accepted 15 July 2004}

\abstract{
The time evolution of the spectral index and the non-thermal flux in 24
impulsive solar hard X-ray flares of GOES class M was studied in RHESSI 
observations. The high spectral resolution allows for a clean separation
of thermal and non-thermal components in the 10--30 keV range, where most
of the non-thermal photons are emitted. Spectral index and flux can thus
be determined with much better accuracy than before.
The spectral soft-hard-soft behavior in rise-peak-decay phases is discovered
not only in the general flare development, but even more pronounced in
subpeaks.
An empirically found power-law dependence between the spectral index and the
normalization of the non-thermal flux holds during the rise and decay phases
of the emission peaks. It is still present in the combined set of all flares.
We find an asymmetry in this dependence between rise and decay phases of
the non-thermal emission.
There is no delay between flux peak and spectral index minimum.
The soft-hard-soft behavior appears to be an intrinsic signature
of the elementary electron acceleration process.
\keywords{Sun: flares -- Sun: X-rays, gamma rays -- Acceleration of particles}
}

\maketitle

%
\section{Introduction}
%

Non-thermal hard X-ray emission during impulsive solar flares is highly
variable, often showing activity peaks and dips with durations
ranging from seconds up to several minutes.
This behavior can be observed in the
largest X class flares as well as in smaller B and C class flares.
It was early recognized (Parks \& Winckler \cite{parks69}; Kane \& Anderson
\cite{kane70}) that the hardness of the photon spectrum can also change with
time, and, furthermore, that there is a direct correlation between the
hard X-ray flux and the spectral hardness.
Since this implies that the flare spectrum starts soft, gets
harder as the flux rises and softer again after the peak time, the term
\emph{soft-hard-soft\/} (SHS) was coined to describe this behavior.
Later observations of major flares (Benz \cite{benz77};
Brown \& Loran \cite{brown85}; Lin \& Schwartz \cite{lin86};
Fletcher \& Hudson \cite{fletcher02}; Hudson \& F{\' a}rn{\'{\i}}k
\cite{hudson02}) confirmed the SHS pattern. 
However, flares were also observed that systematically hardened with time
(Frost and Dennis \cite{frost71}; Cliver et al.
\cite{cliver86}; Kiplinger \cite{kiplinger95}),
thus showing a \emph{soft-hard-harder\/} (SHH) pattern.
Current wisdom suggests that SHH flares represent gradual,
long duration events. These are much less frequent than impulsive events.

The non-thermal photons usually follow a power-law distribution
in energy. The power-law index, $\gamma$, can be
directly related to the energy distribution of the electron flux impinging
on the target (assuming a model for the bremsstrahlung emission),
and implicitly to the acceleration process. The evolution of spectral
index and flux reflects a development in the accelerator. Thus, the relation
between index and flux is an observational constraint for acceleration
theories.

While it seems to be well established that impulsive flares have an SHS
spectral dynamic, much less is known about the \emph{quantitative} relation,
if any, between the photon spectral index and
the non-thermal X-ray flux during the burst.
The Reuven Ramaty High-Energy Solar Spectroscopic Imager (RHESSI)
spacecraft (Lin et al. \cite{lin02}), which observes hard
X-rays and $\gamma$-rays from the sun, is ideally suited to explore
this relation. Its key features of high spectral resolution (1 keV in the
X-ray range) and coverage of the low-energy range (down to 3 keV)
allow us to separate the thermal continuum from the non-thermal component
of the spectrum, to study also the 10-30 keV region where most of the
non-thermal photons are emitted, to identify and account for peculiar
spectral features (like breaks in the power-law etc.),
and to follow the evolution of the non-thermal part right from the
onset of the flare. Therefore, the spectral index and flux of RHESSI
non-thermal photons can be studied with much higher precision than
previously.

This paper presents a large set of measurements of the non-thermal component
from 24 different solar flares and investigates quantitatively the
relation between the non-thermal flux and the spectral index.
The flares were selected in such a way as to represent
the class of impulsive flares with strong non-thermal emission.
In Section \ref{method} we give a detailed description of the selection
and data reduction process that yields the dataset which is then
analyzed in Section \ref{gammaf35} and discussed in Section \ref{discussion}.

%
\section{Observations and Data Reduction}
\label{method}
%

Our main observational goal is the accurate study of the time evolution
of the spectral index $\gamma$ and the non-thermal X-ray flux in a
representative sample of solar flares, using data from RHESSI.
In this section we give a detailed account of the different steps that were
undertaken in the data analysis process, starting from the event selection.

\subsection{Event selection}

The event selection has to be very careful in order to pick a representative
collection of flares. Ideally, one would analyze all the observed events or a
randomly chosen subset thereof.
In practice, however, instrumental issues reduce the
freedom of choice, since not all the events are equally suitable for the
different tasks of high precision spectral analysis.
A detailed discussion about the
RHESSI onboard detectors and their use for spectroscopy can be found in
Smith et. al. (\cite{smith02}). 
We limited our analysis to flares having a peak soft X-ray
flux larger than GOES class M1 and smaller than X1.
These have fairly large count rates, but are not too heavily affected by
pulse pileup.
176~M-class flares were reported in the RHESSI flare list in the period from
13~February 2002 to 31~November 2002.
From this collection we restricted our analysis to the 79 flares observed
with a constant attenuator state of 1 (thin attenuator in) and no
front-segment decimation.
Therefore we do not need to deal with attenuator motions
and decimation state changes during the flare, and we have the best
conditions for spectroscopy of M-class flares.

From this selection of 79 flares we dropped the ones which had no emission
above the background in the 25--50 keV band, as determined by visual inspection
of the observing summary light curves. Since we expect the bulk of the thermal
radiation from M-class flares to be emitted mostly below 25~keV, this
condition introduces a bias toward flares with substantial
non-thermal emission.
This is not a severe restriction, since we want to study specifically the
non-thermal emission, and it would be very hard to ascertain the properties
of any weak non-thermal emission anyway. 
In order to have enough data for meaningful time series, we additionally
required the peak in the 25--50 keV band being more than 3~minutes away from
any interruption in the data, as caused by the spacecraft entering or
leaving the shadow of the Earth, the South Atlantic Anomaly (SAA), etc.
We also dropped the events in which charged particle precipitation 
significantly increased the background counts
during the time of enhanced emission in the 25--50 keV band.
These additional criteria dropped the number of events to 32.

%
\subsection{Data reduction and analysis}

For each event in the list, we determined:
\begin{itemize}
\item
 A contiguous time interval of sunlight containing the event without data gaps.
\item
 The peak time of the observing summary count light curve in the
energy band 25--50~keV.
\item 
 The RHESSI rotation period at peak time.
\item
 The location of the source on the sun.
\end{itemize}
The flare locations were taken from the automatically computed
positions given by the RHESSI Experimental Data Center (HEDC)
(Saint-Hilaire et. al. \cite{saint-hilaire02}). Visual inspection of the
corresponding images confirmed that in all cases correct and accurate positions
were given, with the exception of 3 flares for which the RHESSI aspect solution
(Fivian et. al. \cite{fivian02}; Hurford \& Curtis \cite{hurford02}) quality
was insufficient to permit reconstruction of meaningful images.
We then proceeded to generate RHESSI count spectrograms for each flare in the
uninterrupted sunlight time interval, with a time binning equal to the
spin period at peak time (which in all cases is very close to 4 s)
and an energy binning
of 1~keV from 3 to 6~keV, 0.33~keV from 6 to 13~keV, 1~keV from 
13 to 36~keV, 2~keV from 36 to 60~keV, 5~keV from 60 to 120~keV,
10~kev from 120 to 200~keV, 20~keV from 200 to 300~keV.
Pileup correction was enabled with a deadtime threshold of 5\%.
We only used the front segments of the detectors, and
systematically excluded the detectors 2 and 7, which have lower energy
resolution. For some flares we also excluded detector 8 (which does not
deliver good data when the onboard transmitter is active). The full spectral
response matrix (SRM) was computed 
for each spectrogram, using the HEDC flare positions to enable position
dependent corrections for the flares whose position was known.
We discarded 2 events for which we were unable to generate the SRM. 

\begin{figure}
\resizebox{\hsize}{!}{\includegraphics{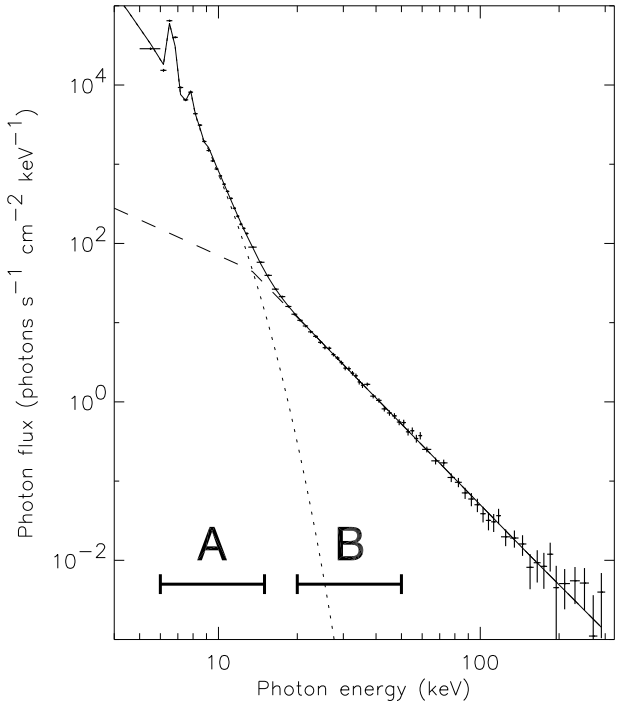}}
\caption{A RHESSI photon spectrum for 9 November 2002 at 13:14:16 UT,
integrated over one rotation period of approximately 4 s.
Overlaid on it, an isothermal thick-target bremsstrahlung emission (dotted
line) with temperature~$T=16.7\;\mathrm{MK}$ and emission measure
$\emes =7.54\!\cdot\!10^{48}\;\mathrm{cm}^{-3}$, a power-law
(dashed line) with spectral index $\gamma=3.39$ and normalization
$\flux{50}=0.525$, and a low-energy turnover $\eturn =13.4\;\mathrm{keV}$.
The continuous line represents the sum of these two components.}
\label{model_spectrum}
\end{figure}

To derive the spectral indices for the spectra, we use the forward fitting
method implemented by the SPEX code (Schwartz \cite{schwartz96};
Smith et al. \cite{smith02}).
The procedure requires the user to choose a model photon spectrum, which is 
multiplied with the instrument response matrix and then fitted to the
observed count spectrum.
The best-fit parameters are given as output. To obtain the time evolution
of the parameters, the fitting procedure is performed for each time bin
in the spectrogram of the time interval of interest.
We have chosen to use a photon spectral model featuring a power-law with
a low-energy turnover in addition to a thermal thick-target bremsstrahlung
emission.
The negative power-law index below the low-energy turnover was fixed at 1.5.
Hence there are 5 free parameters in the model: the temperature~$T$ of the
assumed isothermal emission and its emission measure~$\emes$; for the
non-thermal component the power-law index~$\gamma$,
the normalization of the power-law $\flux{\eref}$ at the
(fixed) normalization energy $\eref$ and the low-energy turnover~$\eturn$.
The non-thermal part of the spectrum is thus given by
\begin{equation}
F(E) =
\left\{
   \begin{array}{ll}
      \displaystyle{\flux{\eref} \left(\frac{E}{\eref}\right)^{-\gamma}}
   &  E>\eturn \\
      \displaystyle{\flux{\eref} \left(\frac{\eturn}{\eref}\right )^{-\gamma}
               \left(\frac{E}{\eturn}\right)^{-1.5}}
   &  E<\eturn
   \end{array}
\right.
\end{equation}
An example photon spectrum with an overlay of the best-fit model is shown
in Fig.~\ref{model_spectrum}.
%
%
For all the events we selected background time intervals (preferentially before
and after the flare) in each of the following 4 energy bands:
3--12, 12--30, 30--60, 60--300 keV. The counts in the different background
intervals were then fitted to a polynomial of degree varying from 0 to 3,
which was used to interpolate the background intensity during the event.
We defined then the \emph{fitting time interval\/} as the time when the
emission in the band 30--60 keV was significantly above the background level.

\subsection{Automatic spectral fitting}

A time dependent determination of the model's best-fit parameters for
30 flares lasting a few minutes with 4 s data bins
requires more than thousand fittings.
To reduce the burden of the work involved in the data analysis,
we implemented an automatic fitting procedure. However, automatic
procedures have their own drawbacks, in particular if the fitting happens
to converge towards a wrong local minimum of~$\chi^2$, sometimes
giving as a result spectacularly wrong fittings.
We decided to settle for the following compromise:
we let the fittings be computed automatically, but we
visually inspected the results afterwards, and eliminated then the
obviously wrong ones without making any attempt to recompute them.
As a matter of fact, we iterated the procedure described above
for a few rounds, each time improving the automatic fitting routine.
After the last run we had to discard only a few fittings
at the beginning and at the end of some fitting time intervals.
Here follows a basic description of the algorithm used by the automatic
fitting routine:
\begin{enumerate}
\item\label{therm_fit}
 A fitting of~$T$ and~$\emes$ for an isothermal emission is made in the
 low-energy range A shown in Fig. \ref{model_spectrum},
 where one can safely assume that the thermal emission is  dominant.
\item\label{nonthermrange} 
 The range for the initial fit of the non-thermal part is defined as B in
 Fig.~\ref{model_spectrum}. If the emission does not exceed   
 0.1 photons $\mathrm{s}^{-1}$ $\mathrm{cm}^{-2}$ $\mathrm{keV}^{-1}$
 or five times the thermal emission as found in step \ref{therm_fit},
 the range of the fit is reduced to the interval where it does satisfy
 these conditions.
\item\label{nonth_fit}
 A photon spectral model with a fixed thermal emission with~$T$ and~$\emes$
 as found in step~\ref{therm_fit} and a (not broken) power-law is fitted in
 the non-thermal range defined in step~\ref{nonthermrange}.
\item\label{free_fit}
 The parameters resulting from step~\ref{therm_fit} and~\ref{nonth_fit}
 are used as initial estimates for the final fitting of all the~5 free
 parameters of the model to the spectrum in the energy range 6--150~keV.
 The low-energy turnover initial estimate is taken as either the fitted
 $\eturn$ from the previous spectrum in the time sequence, or a default value. 
\end{enumerate}
We emphasize that the first steps described above only provide the initial
estimates for the parameters, which are then let \emph{totally free\/} in the
last fitting over the whole energy range. Steps \ref{therm_fit}--\ref{nonth_fit} simply attempt to give
reasonable initial guesses for the parameters. The energy ranges A and B
and the default values used by the routine are fixed and were determined
empirically to achieve the largest possible number of good fittings.
We have experimented with alternative spectral models. They included for
example models with a slightly hotter isothermal component,
or two components of thermal emission, but no non-thermal emission.
Such models fit the observed spectra with similar values of~$\chi^2$,
when the power-law index $\gamma$ is greater than about~8.

\subsection{Best-fit parameters selection and results}
\label{finalsel}

The automatic fitting routine failed to provide results for~4 events.
Its output consists of the best-fit parameters for a total of 1\,566 spectral
fittings from~26 events.
The visual inspection of all the fittings allowed us
to eliminate the spectra which were badly fitted because they would
have required a broken power-law model. We also chose to discard all the
spectra whose power-law component had an index similar to the logarithmic
derivative of the thermal emission around the energy at which it
was only about as strong as the background, because in such a
case it is very difficult to ascertain the reality of any non-thermal
emission.
After the selection process explained above, we were left with a total of~911
fittings for~24 events, spanning a total time of 3\,722~s.
The number of fittings for each event ranges from~5 to~212, with an average
of~38. The~24 events with at least 5 good fittings are our final selection.
They are listed in Table \ref{eventlist}. Each fit consists of the
5 parameters $T$, $\emes$, $\gamma$, $\flux{0}$ and $\eturn$.
All events are relatively short, have often many peaks and comply with the
definition of impulsive flares.

%
%
\begin{table}
\centering
\caption{List of the selected 24 events.
Peak flux means the fitted non-thermal flux at 35 keV at peak time in
photons $\mathrm{s}^{-1}$ $\mathrm{cm}^{-2}$ $\mathrm{keV}^{-1}$.
The peak time given reflects the time of maximum flux \emph{after\/}
the fitting selection, and may therefore not coincide with the peak of a
light curve at 35 keV.}
\begin{tabular}{cccccc}
\hline\hline
Event    & Event date  & Peak & Peak & Peak     & Number \\ 
Nr.      &             & time & flux & $\gamma$ & of fittings \\
\hline
1& 20 Feb 2002 & 09:54:04 & 0.315 & 6.8 & 54  \\  
2& 20 Feb 2002 & 16:22:58 & 0.168 & 3.8 & 11  \\  
3& 20 Feb 2002 & 21:06:05 & 2.388 & 4.1 & 23  \\  
4& 25 Feb 2002 & 02:56:42 & 0.390 & 5.5 & 29  \\  
5& 26 Feb 2002 & 10:26:52 & 4.915 & 3.2 & 20  \\  
6& 15 Mar 2002 & 22:23:06 & 0.243 & 4.9 & 125 \\  
7& 04 Apr 2002 & 10:43:55 & 0.261 & 5.2 & 49  \\  
8& 04 Apr 2002 & 15:29:16 & 2.124 & 4.7 & 32  \\  
9& 09 Apr 2002 & 12:59:51 & 0.403 & 4.9 & 41  \\  
10& 14 Apr 2002 & 03:24:44 & 0.715 & 4.6 & 13 \\  
11& 17 Apr 2002 & 00:38:34 & 0.253 & 4.4 & 8  \\  
12& 24 Apr 2002 & 21:50:23 & 0.712 & 4.0 & 16 \\  
13& 01 Jun 2002 & 03:53:41 & 2.473 & 3.0 & 25 \\  
14& 16 Aug 2002 & 22:10:30 & 1.629 & 4.9 & 18 \\  
15& 17 Aug 2002 & 01:02:04 & 0.076 & 4.3 & 5  \\  
16& 23 Aug 2002 & 11:59:05 & 0.220 & 6.4 & 14 \\  
17& 24 Aug 2002 & 05:43:23 & 0.216 & 5.7 & 84 \\  
18& 27 Aug 2002 & 12:28:38 & 1.617 & 3.1 & 8  \\  
19& 29 Sep 2002 & 06:36:18 & 4.261 & 3.8 & 30 \\  
20& 29 Sep 2002 & 14:46:43 & 0.405 & 3.9 & 40 \\  
21& 30 Sep 2002 & 01:48:25 & 0.107 & 7.3 & 18 \\  
22& 04 Oct 2002 & 00:41:13 & 1.109 & 4.6 & 20 \\  
23& 09 Nov 2002 & 13:16:36 & 5.776 & 3.2 & 212\\  
24& 14 Nov 2002 & 22:24:40 & 1.978 & 3.8 & 16 \\ \hline 
\end{tabular}
\label{eventlist}
\end{table}
%
%

Fitting free parameters to data may introduce hidden dependencies between
them. It is important for the study of the index-flux relation to assess
the effect of the fitting procedure.
For this purpose we have compared the index-flux relation before and after
fitting.
For the flare of 9 November 2002 (the one with the longest time series)
we computed two supplementary time series for the flux and spectral index
from the uncalibrated count-rates total flux in the energy bands 26--35 keV
and 35--44 keV and from their ratio.
This is a much simpler and cruder way of determining the values of the
two parameters that does not require fitting. Although
the absolute values of the parameters will differ, they preserve the temporal
variations.
%
%
%
%
%
%
\begin{figure*}
\resizebox{\hsize}{!}{\includegraphics{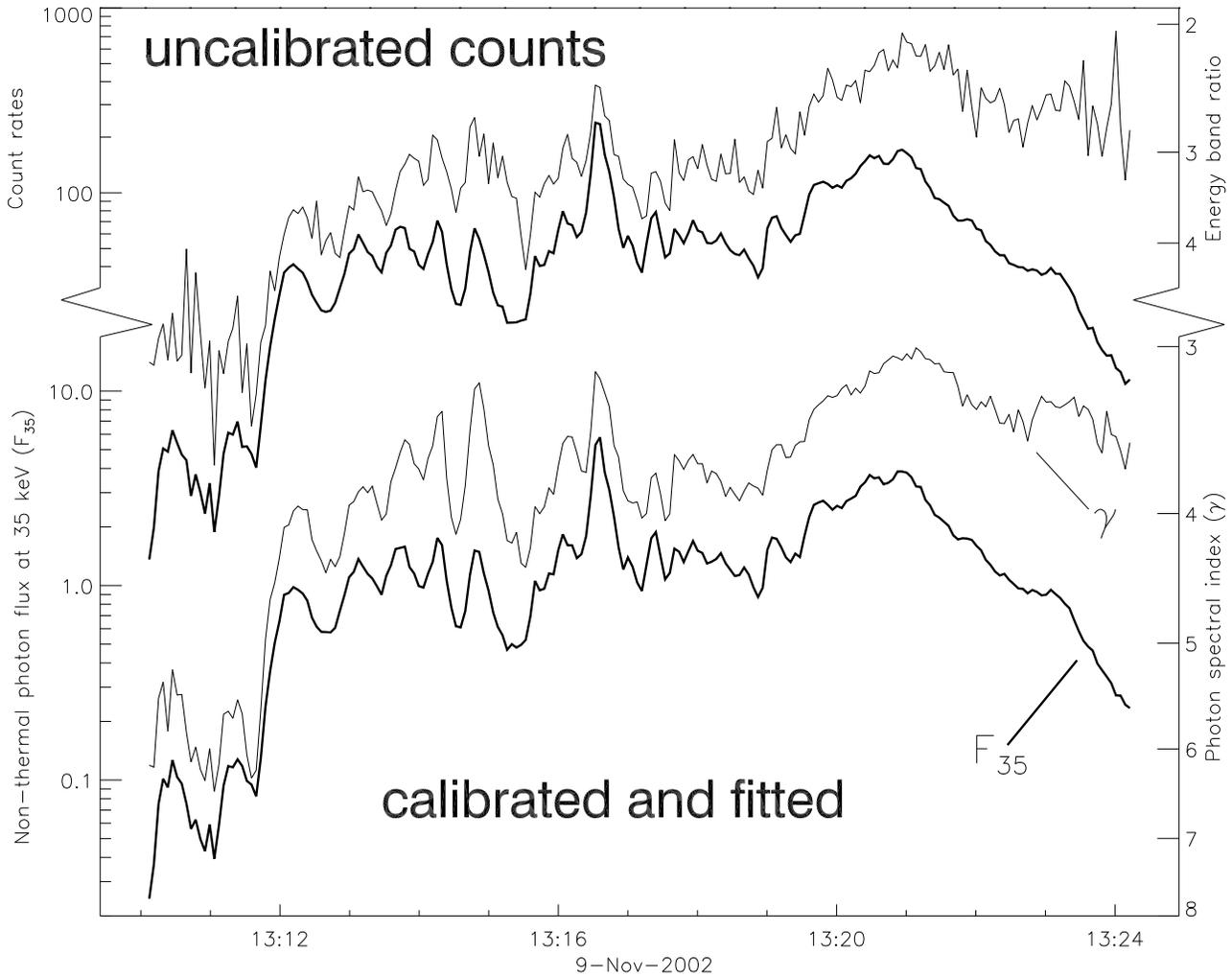}}
\caption{%
{\it Top\/}: Spectral index (thin line) and flux (thick line)
obtained from the uncalibrated total count rates
flux in the energy bands 26--35 keV and 35--44 keV and their ratio.
{\it Bottom\/}: Spectral index $\gamma$ (thin line) and non-thermal flux
$\flux{35}$ at 35 keV in photons
$\mathrm{s}^{-1}\mathrm{cm}^{-2}\mathrm{keV}^{-1}$
(thick line) for the event of 9 November 2002, obtained by spectral fitting.}
\label{time_ev_30}
\end{figure*}
Figure \ref{time_ev_30} shows an extremely close similarity between
the uncalibrated count rates in the 26--44 keV energy range and the fitted
flux. It proves that the fitting preserves the time evolution of the
observed counts with high precision.
The spectral ratio derived from uncalibrated counts in the two energy
bands is normalized to the energy ranges, but differs from the fitted
$\gamma$ due to the lack of calibration. It is significantly noisier
than $\gamma$
because of the broader energy range used for the fitting, which
includes non-thermal photons down to the thermal cross-over near
15 keV.
Yet the two spectral parameters follow each other
extremely well in time. This confirms that the dependence introduced
by fitting is negligible.

To further check that no cross-talk between thermal and non-thermal
parameters is introduced by the fitting procedure, we smoothed the
curve representing the temperature $T$ as a function of time for flare~23
using a smoothing filter with a time window of 180~s,
and recomputed all the fittings forcing the temperature to follow
the smoothed curve. No significant differences were found between the new
values obtained for the spectral indices and non-thermal fluxes and the old
ones. This shows that any short-term variation of the fitted temperature
during an emission peak do not significantly influence the behavior of
$\gamma$ and $\flux{35}$.

%
%
\section{Relation between flux and spectral index}
\label{gammaf35}
%
%
In this section we will investigate the relation between the spectral index
and the non-thermal flux in a quantitative way.
For the non-thermal flux there are two options: the strength of the
non-thermal flux $\flux{E_0}$ at a chosen, fixed normalization energy
$\eref$ or the total non-thermal flux $\ftot=\int F(E)\,{\mathrm d}E$.
The first option has the disadvantage of requiring a supplementary arbitrary
parameter $E_0$.
Therefore, it would be conceptually preferable to choose the second.
However, the non-thermal flux is only observed in the energy range where it
is larger than both the background and the thermal emission.
Hence it is not straightforward to compute $\ftot$, because we do not know
the non-thermal emission outside the observed range.
The main uncertainty in $\ftot$ comes from errors in the
estimates of the low-energy turnover $\eturn$,
especially for steep spectra, where small errors in $\eturn$ 
can produce large changes in $\ftot$.
Since spectral fittings are poorly suited to accurately determine
$\eturn$ and hence $\ftot$ (for a detailed discussion,
see Saint-Hilaire and Benz, \cite{saint-hilaire04}), the large uncertainties
in $\ftot$ rule out its use as a main parameter for our study.
We are left with the first option, but we have to choose $\eref$.
It should lie in an energy range where the non-thermal emission is actually
observed and fitted, to compare observable quantities.
The non-thermal emission is best observed and identified in the energy range
20--50~keV, and hence we choose the center $\eref= 35\,\mathrm{keV}$, and
investigate the $\gamma$-$\flux{35}$ relation.
In Section \ref{discussion} we will consider 
the implications of changes in $\eref$ from the chosen 35 keV.

\begin{figure}
\resizebox{\hsize}{!}{\includegraphics{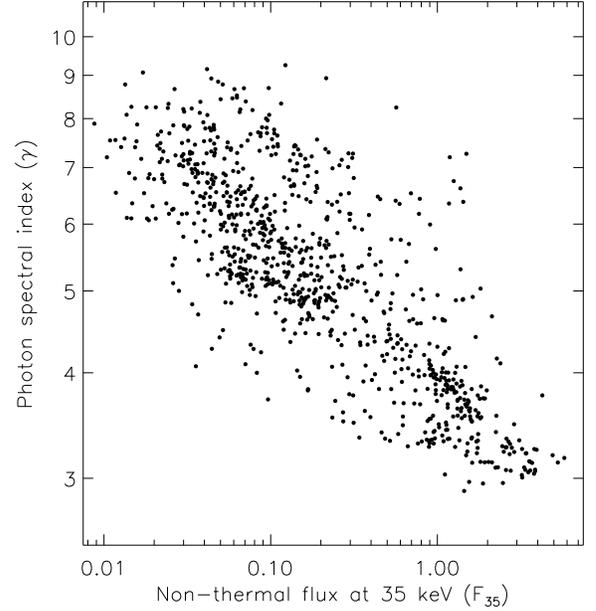}}
\caption{Plot of $\gamma$ versus the fitted non-thermal flux at 35 keV
(given in photons $\mathrm{s}^{-1}$ $\mathrm{cm}^{-2}$ $\mathrm{keV}^{-1}$).
All the 911 data points from the 24 events are shown.}
\label{gammaspindextot}
\end{figure}

Fig. \ref{gammaspindextot} presents a logarithmic plot of $\gamma$ versus
$\flux{35}$ for all of the 911 data points.
%
%
The plot clearly shows the overall SHS trend. 
The cross-correlation coefficient of $\ln\gamma$ versus $\ln \flux{35}$
is $r=-0.80\,\pm\,0.03$,
where the uncertainties given represent the 99\% confidence range.
The cross correlation coefficient $r$ is significantly lower than 1 because of
the scatter in the data, which is real and not due to measurement errors.
The $\gamma$ vs. $\flux{35}$ relation can be approximated by a power-law
model
\begin{eqnarray}
\gamma     &=& A \flux{35}{}^{-\alpha} \quad\mathrm{or, equivalently,}
\label{gammafluxfunc}\\
\ln\gamma &=& \ln(A) - \alpha \ln \flux{35}\,.
\label{gammafluxfunclog}
\end{eqnarray}
In Eq. (\ref{gammafluxfunc}) and in the following, $\flux{35}$ is used
as a dimensionless number given by the normalization factor at 35 keV
divided by the unit flux (1 photon $\mathrm{s}^{-1}$ $\mathrm{cm}^{-2}$
$\mathrm{keV}^{-1}$).
The constants $A$ and $\alpha$ can be obtained by means of a linear
least-squares (LS) regression of the quantities $\ln\gamma$ vs.
$\ln \flux{35}$.
However we note that:
\begin{enumerate}
\item The $\gamma$-$\flux{35}$ data have a fairly large scatter.
\item We have to treat the variables $\gamma$ and $\flux{35}$ symmetrically,
since it is not \emph{a priori\/} clear that one of them is a function of the
other, and thus there is no reason to take either one as a \emph{dependent\/}
variable.
\end{enumerate}
In such a case, Isobe et. al. (\cite{isobe90}) suggest that the best fitting
parameters are obtained by the LS bisector method.
The method consists of taking the line that bisects
the LS($\ln\gamma\,|\,\ln\flux{35}$) and LS($\ln\flux{35}\,|\,\ln\gamma$)
regression lines, where LS($\,y\,|\,x$) means the least-square regression
of the dependent variable $y$ against the independent variable $x$.
The LS bisector gives for our model parameters: $A = 4.043\pm 0.032$
and $\alpha=0.197\pm 0.003$.
Computing $\alpha$ by LS($\ln\gamma\,|\,\ln\flux{35}$) we get
$\alpha=0.154\pm 0.003$, and using  LS($\ln\flux{35}\,|\,\ln\gamma$)
$\alpha=0.241\pm 0.006$.
The latter two values define a
confidence range for $\alpha$, such that $\alpha=0.20\pm 0.05$.
These values will be refined in the following. They may be used for
a future comparison, e.g. with a peak-flux analysis.

%
%

\begin{figure*}
\resizebox{\hsize}{!}{\includegraphics{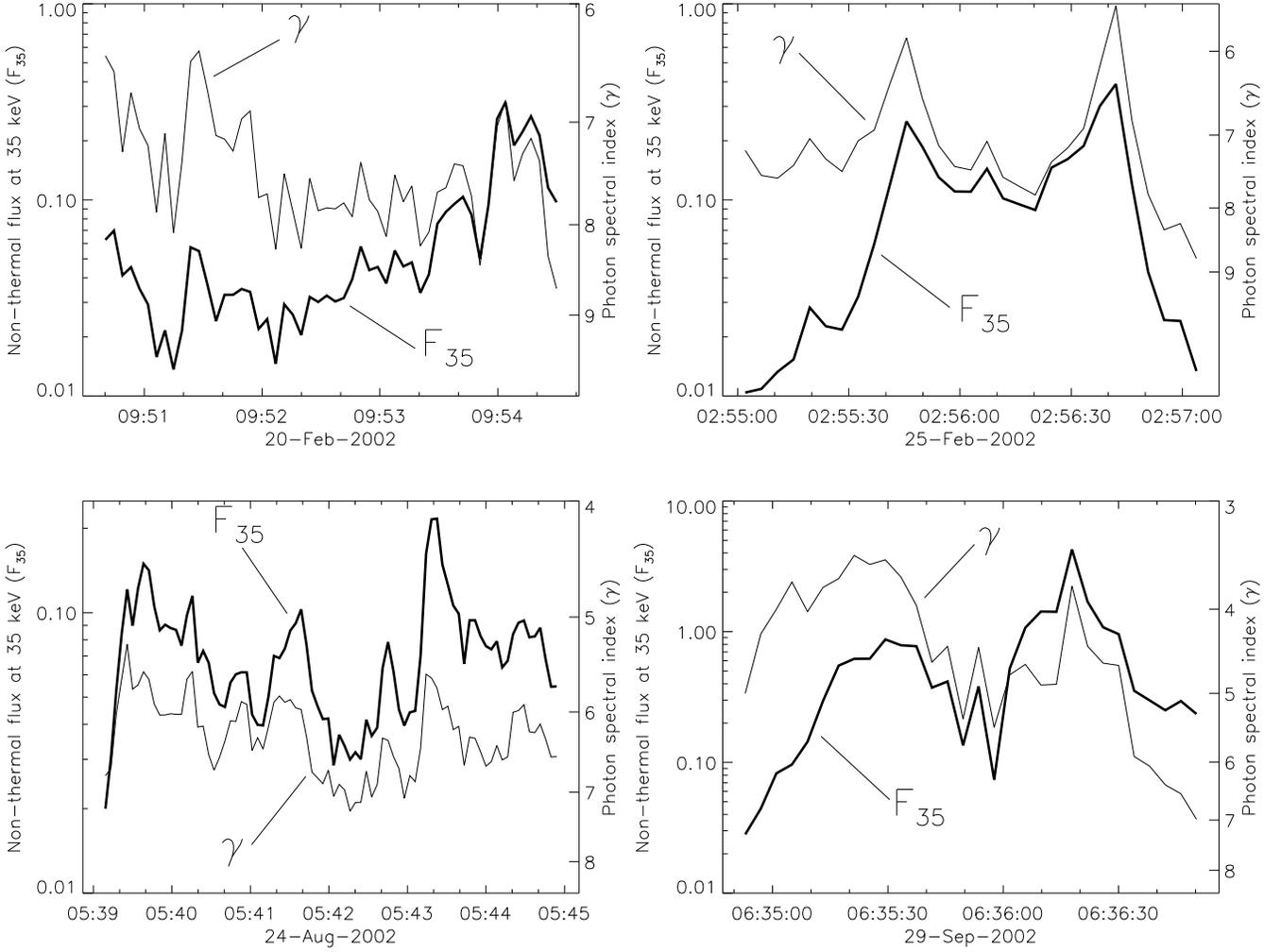}}
\caption{Time evolution of the photon spectral index $\gamma$ (thin line)
and the non-thermal flux at 35 keV $\flux{35}$ (thick line) for 4 flares.}
\label{timevol}
\end{figure*}

%
%
The overall behavior of the plot in Fig. \ref{gammaspindextot} results from
the superposition of points from different flares.
We now want to look in some more detail at the behavior of single flares.
Fig. \ref{timevol} gives the time evolution of $\gamma$ and $F_{35}$ for
4 flares. 
Anti-correlation of $\flux{35}$ and $\gamma$  can clearly be seen
in all of them.
However, it is evident from a close inspection of the light curves
that a model of a strict functional dependence like the one of Eq.
(\ref{gammafluxfunc}) will not work well for an entire flare,
since there are consecutive peaks in $\flux{35}$ with about the same
height, but having different minimum values of $\gamma$.
Nevertheless each peak shows an unmistakable SHS pattern.
It seems likely that the proposed power-law model suits better
the behavior of single peaks than whole flares.
To check this, we computed the vertical scatter
of the data points around their LS($\ln\gamma | \ln\flux{35}$) regression
line for the set of all the data points, for the 24 subsets of points
belonging to each flare, and for 141 subsets of points belonging to
70 rise and 71 decay phases of peaks during the flares. The rise
and decay phases were selected with the requirement that each phase
consists of at least 3 consecutive data points.
The vertical scatter $\sigma_I$ of a subset $I$ of $n$ points
$(\ln\flux{35}^i\, , \ln\gamma^i)\, ,\,i\in I$ around the straight line 
defined by Eq. (\ref{gammafluxfunclog}) with
LS($\ln\gamma|\ln\flux{35}$) parameters slope $\alpha_I$ and intercept
$\ln A_I$ can be computed by
\begin{equation}
\sigma^2_I=\frac{1}{n-2}\sum_{i=1}^n \left(\ln\gamma^i-\ln A_I+
\alpha_I\ln\flux{35}^i\right)^2,
\end{equation}
where $n-2$ are the degrees of freedom of the subset with
the fitted straight line. 
For the set of all 911 points $\sigma=0.165$.
The average of $\sigma$ computed for each single flare is $0.094$.
The average of $\sigma$ for each rise and decay phase is $0.020$ and,
respectively, $0.017$.
The improvement in sigma going from the total dataset
to the flares and then to the rise/decay phases
shows that indeed single peaks are better represented by the model,
in the sense that they have much less vertical scatter around
the regression line.

\begin{figure*}
\resizebox{\hsize}{!}{\includegraphics{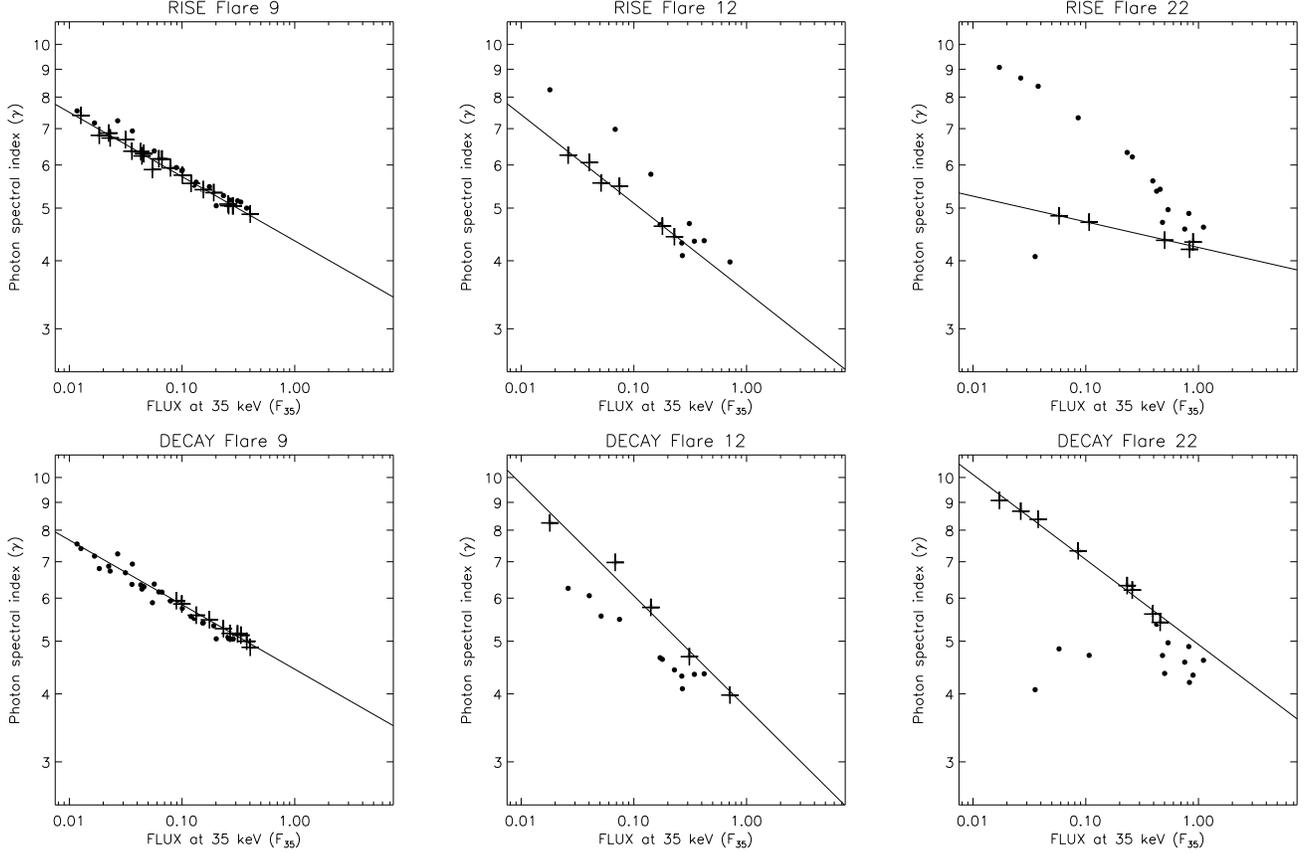}}
\caption{Spectral index $\gamma$ versus flux at 35 keV $\flux{35}$ for 
3 flares.  The plus signs in the top and bottom row of plots mark the points
forming the longest uninterrupted rise and decay phase in the flare.
The regression line of the selected data points is shown.}
\label{peakfit}
\end{figure*}

Fig. \ref{peakfit} 
shows the $\gamma$ vs. $\flux{35}$ plot for three events.
For each flare, the points belonging to the longest
rise (top row) and decay (bottom row) phase are represented by plusses
and their LS($\ln\gamma\,|\,\ln\flux{35}$) regression line is drawn.
Since here we have much less scatter, there is little difference between
the different LS regression schemes.
It can be seen that the scatter around the fitted line is low
during the rise/decay phases.
Interestingly, in some flares the fitted line is steeper in the decay phases.

\begin{figure}
\resizebox{\hsize}{!}{\includegraphics{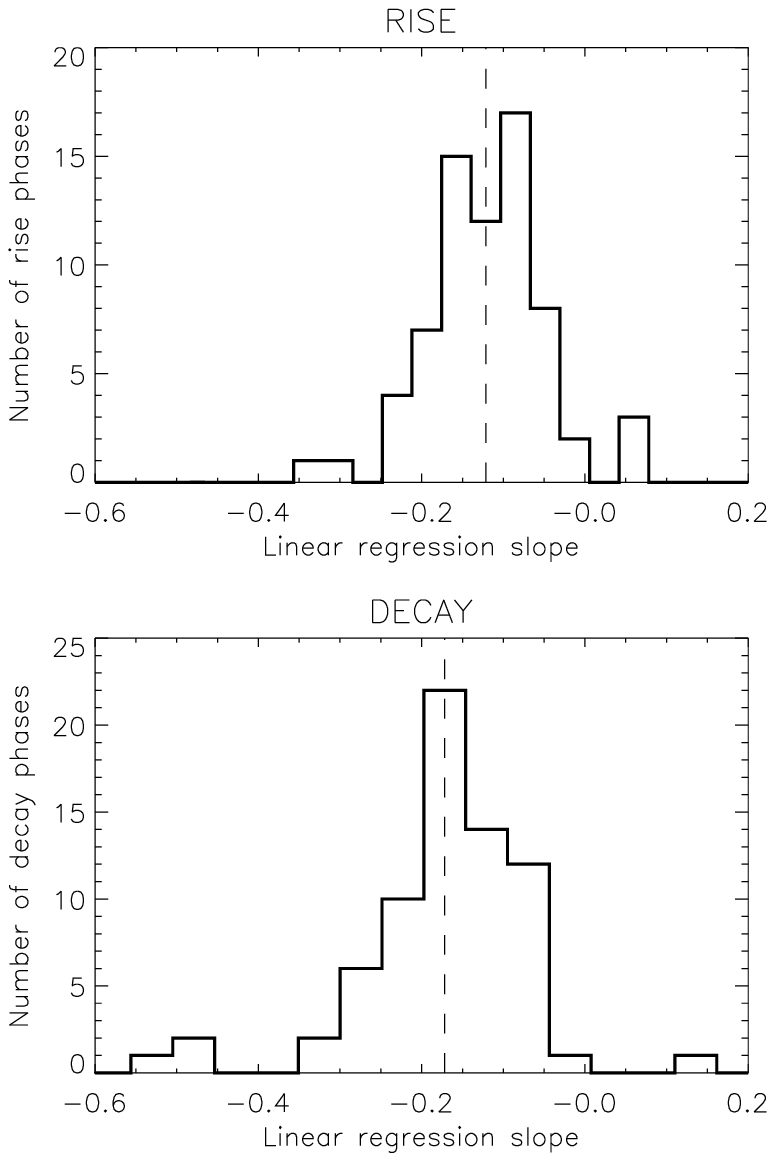}}
\caption{Distribution of the linear regression slopes for the rise and
decay phases. The average value is marked by the dashed line.}
\label{histolines}
\end{figure}

We show the distribution of the slope of the fitted lines in
Fig. \ref{histolines}.
The bin width is 0.5 standard deviations of the measured slopes.
The average value for the rise phase is 
\begin{equation}
\alpha_{\rm r}= 0.121\pm 0.009
\end{equation}
and the one for the decay phase is
\begin{equation}
\alpha_{\rm d}= 0.172\pm 0.012.
\end{equation}
The standard deviations are $\sigma_{\rm r}=0.073$ and
$\sigma_{\rm d}=0.103$, respectively.

The difference in averages is about 5 times the standard errors of the mean,
and therefore the difference between the two cases is statistically
significant.
We see that for most of the rise phases there is a soft-hard trend
(negative slope) and for most of the decay phases there is a hard-soft
trend (also described by a negative slope). The number of rise phases
with slope smaller than 0.04 are 5 out of 70, and the number of decay
phases with slope smaller than 0.04 are 2 out of 71.
Therefore the SHS behavior is a nearly universal trend in peaks of
non-thermal emission.

We also investigated whether there is a significant delay in the
correlation of $\flux{35}$ and $\gamma$.
We defined the delay as the time of the minimum of the quadratic
interpolation curve going through the 3 cross-correlation coefficients
corresponding to a lag of $-1$, 0 and 1 time bins of about 4 seconds.
The interpolation enhances considerably the time resolution, as the
noise is small. The distribution of the delays is relatively broad,
centered at $-0.32$ s, with a standard deviation of 1.1 s and
with extreme delays up to $\pm 3$ s.
The average of the delays  does not significantly differ from 0 since
the standard error of the mean is $\pm 0.23$ s.

Furthermore, we tried to see if there is any evidence in the data
for the presence of a \emph{pivot point}, i.e. a fixed point with coordinates
$(E^*,F^*)$ common to all the spectra in a rise/decay phase.
We note that such a concept corresponds to a model that does not
yield a power-law dependence of $\flux{35}$ and $\gamma$,
but instead
\begin{equation}
\gamma=- \frac{\ln \left( F^*            / \flux{35} \right)}
	      {\ln \left( 35\mathrm{keV} / E^*       \right)},
\end{equation}
contrary to Eq. (\ref{gammafluxfunc}).
We computed the intersections of the power-law fits to the non-thermal
component of the photon spectrum of all 4~s time bins in each
rise phase of every flare with all other non-thermal components in the
same phase.
While this procedure is quite
sensitive to errors for nearly parallel lines, the total distribution
yields a clue whether there is any virtue in the idea of a pivot point.
The energy distribution of the intersections
peaks around 9 keV, and was larger than half of the maximum
in the energy range 6.5--12.5 keV, but it had comparatively large tails
to very low and very high energies.
There seems to be no real pivot point, but the region of intersection
is relatively narrow.
Therefore, when trying to visualize the time evolution of the non-thermal
spectrum during the rise phase, one does not too badly by imaging
the spectrum as fixed at an energy around 10 keV.
In the course of the flare the non-thermal spectrum rises its high-energy
tail until peak time, and it lowers it again afterwards.


%
\section{Discussion}
\label{discussion}
%

The power-law model for the $\gamma$-$\flux{35}$ relation is admittedly very
simple, yet it provides a good empirical description of the observed
quantities.
The range of validity of the model is limited at very high flux
values, since $\gamma$ has a theoretical lower limit at roughly 1.4,
given by the bremsstrahlung of a monoenergetic beam.
Its major disadvantage is the arbitrary assumption of a
normalization energy, here 35 keV. How do the results shown in Section
\ref{gammaf35} depend on the choice $\eref=35\,\mathrm{keV}$?

Let $E_1$ and $E_2=E_1+\Delta E$ be two normalization energies.
The normalization coefficients $\flux{E_0}$ and $\flux{E_1}$ satisfy
\begin{equation}
\flux{2}=\flux{1}\left(\frac{E_2}{E_1}\right)^{-\gamma}.
\label{F1vsF2}
\end{equation}
Now let us assume that the relation $\gamma=A_1 \flux{1}^{-\alpha_1}$
holds. Using Eq. (\ref{F1vsF2}), this can be written as
\begin{equation}
\label{eq7}
\ln\gamma=\ln A_1-\alpha_1\ln\flux{2}-\alpha_1\gamma\ln\frac{E_2}{E_1}.
\end{equation}
If $\Delta E / E_1 \ll 1$, we can expand the logarithm and get
\begin{equation}
\ln\gamma=\ln A_1- \alpha_1\ln\flux{2}-\alpha_1\gamma\frac{\Delta E}{E_1}.
\end{equation}
The logarithmic derivative of the this expression yields
\begin{equation}
\label{finalgammatransform}
\frac{\mathrm{d} \ln\gamma}{\mathrm{d}\ln \flux{2}}=
\frac{-\alpha_1}{1+\alpha_1\gamma\frac{\Delta E}{E_1}}
\approx -\alpha_1\left(1-\alpha_1\gamma\frac{\Delta E}{E_1}\right).
\end{equation}
The $\ln\gamma$-$\ln\flux{E_2}$ relation is not linear, as $\gamma$ appears on
the right hand side of Eq. (\ref{finalgammatransform}).
If a new exponent $\alpha_2$ were fitted to the $\ln\gamma$-$\ln\flux{E_2}$
data points, we would get according to Eq. (\ref{finalgammatransform}),
\begin{equation}
\alpha_2\simeq \alpha_1\left(1-\alpha_1\gamma\frac{\Delta E}{E_1}\right).
\end{equation}

\begin{figure}
\resizebox{\hsize}{!}{\includegraphics{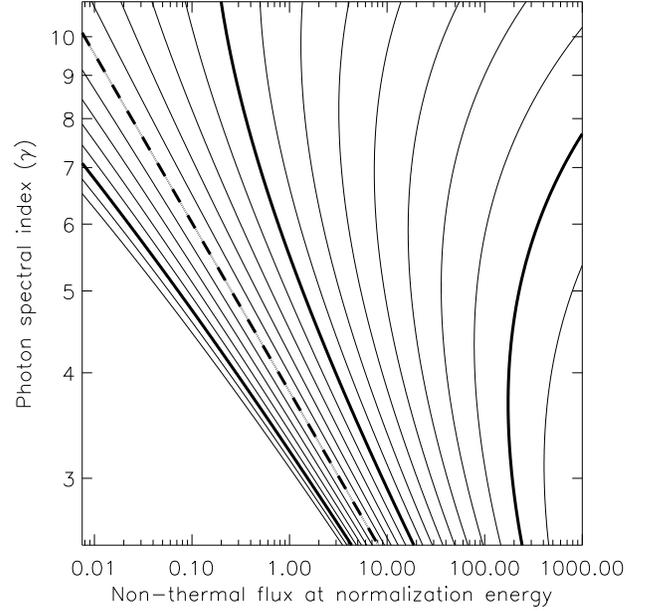}}
\caption{The lines shown here models the relation between the spectral
index $\gamma$ and the non-thermal flux $\flux{E_0}$ at different energies
$E_0$, assuming that $\gamma = A \flux{35}^{-\alpha}$ holds at $E_0=35$ keV
with $\alpha=0.2$ and $A=3.8$.
The dashed line correspond to $E_0=35$ keV, the lower thick line to
$E_0=45$ keV, the upper thick lines to, respectively,
$E_0=25$ keV and $E_0=9$ keV.
The separation between thin lines is 2 keV.}
\label{disceref}
\end{figure}

For $\Delta E \ll E_1$ the relation between $\ln\gamma_2$ and ln$\flux{E_2}$
is still approximately linear but flatter than for $\ln \flux{E_1}$ if
$\Delta E > 0$.
This is confirmed in Fig. \ref{disceref}, where the $\gamma$ vs.
$\flux{E_0}$ relation is shown for different normalization energies $E_0$. 
In Fig. \ref{disceref} the $\gamma$-$\flux{E}$ relation is presented
according to Eq. (\ref{eq7}). In the course of a subpeak the spectral
index and flux move approximately on one of the lines according to the
given normalization energy, assuming the relation (\ref{gammafluxfunc}).
The relation steepens with decreasing normalization energy $E_0$ 
and finally turns over. In a model with a pivot point at energy $E^*$
the $\gamma$-$\flux{E_0}$ relation would be a vertical line for $E=E^*$.

The final goal (not part of this paper) is the comparison of these results with
the prediction of theoretical models for the energy distribution of accelerated
electrons. For this purpose the electron distribution needs to be reconstructed
from the observed photon spectrum. As an example, the analytically solvable
thick target impact model using the non-relativistic Bethe-Heitler cross
section (Brown \cite{brown71}, Tandberg-Hanssen and Emslie
\cite{tandberg-hanssen88}) predicts that an electron power-law distribution
$\Phi(\epsilon) = \Phi_{\epsilon_0} (\epsilon / \epsilon_0)^{-\delta}$
[electrons $\mathrm{s}^{-1}$ $\mathrm{keV}^{-1}$] generates a photon spectrum
$F(E)=\flux{E_0} (E / E_0)^{-\gamma}$
[photons $\mathrm{s}^{-1}$ $\mathrm{cm}^{-2}$ $\mathrm{keV}^{-1}$] with
\begin{equation}
\delta =\gamma + 1
\end{equation}
and
\begin{equation}
\label{equivalent_electron_spectrum}
\Phi_{\epsilon_0}= K\,\flux{E_0}\,E_0^\gamma\,\epsilon_0^{-\delta}\,
\frac{(\delta-1)\,(\delta-2)}{\beta(\delta-2,1/2)},
\end{equation}
where $\beta(x,y)$ is the beta function and the constant $K$ is given by
\begin{equation}
K=\frac{3\pi^2 e^4 \ln \Lambda\,D^2}{Z^2\,\alpha r_e^2\,m_e c^2}
\simeq 6.6\cdot 10^{33}
\,\mathrm{keV}\,\mathrm{cm}^2,
\end{equation}
where $\ln\Lambda$ is the Coulomb logarithm, $D$ the distance from the target
(here 1 AU), $Z$ the average atomic number, $\alpha$ the fine structure
constant and $r_e$ the classical electron radius.
Eq. (\ref{equivalent_electron_spectrum}) indicates that the relation
between electron flux and gamma will not conserve linearity in the log-log
representation. In other words, the assumed linear relation for fitting
photons must be considered as a convenient but rather arbitrary
approximation.

%
\section{Conclusion}
\label{conclusion}
%

This study of the $\gamma$-$\flux{}$ relation in the evolution of the
non-thermal
component of impulsive solar flare hard X-ray emissions exploits the high
spectral resolution of the RHESSI germanium detectors. Contrary to earlier
investigations, the spectral index is not derived from the ratio of a few
channels, but from fitting the spectrum at relatively low non-thermal energies
where most of the photons are emitted. This method eliminates the influence of
the thermal component and improves considerably the noise on the derived
spectral index (Fig. \ref{model_spectrum}). 

The most surprising result of the improved accuracy is the appearance of the
soft-hard-soft behavior on short time scales. The SHS behavior is a feature
seen in nearly all of the non-thermal emission peaks of M-class solar flares.
Whereas SHS was previously considered to be a global property of flares,
Figs. \ref{time_ev_30} and \ref{timevol} 
demonstrate that SHS is a predominantly short-scale phenomenon.
This is the reason why the scatter in the $\gamma$-$\flux{35}$
plot diminishes when rise and decay phases of individual subpeaks
are analyzed separately. 

The novel quantitative analysis of the $\gamma$-$\flux{35}$ relation
has also revealed remarkable properties.
The relation appears linear in double-logarithmic representation
(Fig. \ref{gammaspindextot}). Thus it follows an approximate power law,
$\gamma=A F_{35}^{-\alpha}$. Its average index $\alpha$ is 0.197$ \pm$ 0.003.
The scatter is greatly reduced if individual subpeaks are studied
(Fig. \ref{peakfit}).
In the rise phase of individual flare elements, the average index
$\alpha_r = 0.12\pm 0.01$ is significantly smaller than the index of the
decay phase $\alpha_d = 0.17\pm 0.01$ (Fig. \ref{histolines}).
The path of a subpeak in the $\gamma$-$\flux{35}$ plot (Fig. \ref{peakfit})
follows tendentially a slanted V, with the rise phase forming the
flatter leg.
This amounts to a secondary trend, superimposed on the SHS behaviour,
of a general spectral softening of the non-thermal component in the
course of a subpeak.

The SHS behavior supports the idea that each non-thermal emission peak
represents a distinct acceleration event of the electrons in the flare. The
individual peaks mainly differ by their value for $A$ in the
$\gamma=A F_{35}^{-\alpha}$ relation, presumably due to different physical
parameters in the acceleration region.

It is possible to visualize the $\gamma$-$\flux{35}$ relation by a pivot
point in the
non-thermal spectrum. This point is relatively stable in energy and flux.
The pivot energy was determined as 9 keV in the average with a half-power
distribution of 6.5--12.5 keV. In the course of a peak, the non-thermal
spectrum rises by turning around the pivot point, decreasing $\gamma$ and
increasing the flux beyond the energy of the pivot point. In the decay phase
the spectrum decreases and turns the opposite way. The picture is supported by
the observations of no delay (in the average). We note, however, that the pivot
point model is only an approximation and needs to be further investigated.

The SHS phenomenon of flares, and in particular of subpeaks, contradicts the
idea of the statistical flare in avalanche models (Lu \& Hamilton \cite{lu91}), assuming that each flare and subpeak is composed
of many identical elements that are far below resolution. The superposition of
such subresolution structures in a straightforward avalanche process would not
yield the observed SHS time behavior. 

The subpeaks defined by the SHS behavior thus may be considered as irreducible
flare elements. They have durations of one minute (Fig. \ref{time_ev_30})
to shorter than 8 seconds, the lower limit given by the time resolution
(Fig. 4). The close correlation suggests that there is an intrinsic
dependence between the flux and energy distribution of electrons for any given
elementary acceleration event.
If this is the case, it implies that individual SHS structures cannot
be further resolved, thus form the elementary structures of flares. 

\begin{acknowledgements}
The analysis of RHESSI data at ETH Zurich is partially supported by the Swiss
National Science Foundation (grant nr. 20-67995.02). This work relied on the
RHESSI Experimental Data Center (HEDC) supported by ETH Zurich
(grant TH-W1/99-2). We thank the many people who have contributed to the
successful operation of RHESSI.
We acknowledge the contribution of D. Buser, who has prepared the ground
for this work in his Diploma Thesis (\cite{buser03}).
We thank P. Saint-Hilaire, K. Arzner, S. Krucker, R. Schwartz and H. Hudson
for helpful discussions, and the referee for constructive comments.
\end{acknowledgements}

%

%
\end{document}